**Proof-of-principle demonstration of measurement-device-independent quantum key distribution using polarization qubits**

T. Ferreira da Silva,[1,2,*] D. Vitoreti,[1] G. B. Xavier, [3,4,5], G. C. do Amaral,[1] G. P. Temporão,[1] and J. P. von der Weid[1]

[1]*Centre for Telecommunication Studies, Pontifical Catholic University of Rio de Janeiro, Rua Marquês de São Vicente 225 Gávea, 22451-900 Rio de Janeiro, RJ, Brazil*

[2]*Optical Metrology Division, National Institute of Metrology, Quality and Technology, Av. Nossa Sra. das Graças 50 Xerém, 25250-020 Duque de Caxias, RJ, Brazil*

[3]*Departamento de Ingeniería Eléctrica, Universidad de Concepción, Casilla 160-C, Correo 3, Concepción, Chile*

[4]*Center for Optics and Photonics, Universidad de Concepción, Casilla 4016, Concepción, Chile*

[5]*MSI-Nucleus for Advanced Optics, Universidad de Concepción, Casilla 160-C, Concepción, Chile*

[*]*thiago@opto.cetuc.puc-rio.br*

We perform a proof-of-principle demonstration of the measurement-device-independent quantum key distribution (MDI-QKD) protocol using weak coherent states and polarization-encoded qubits over two optical fiber links of 8.5 km each. Each link was independently stabilized against polarization drifts using a full-polarization control system employing two wavelength-multiplexed control channels. A linear-optics-based polarization Bell-state analyzer was built into the intermediate station, Charlie, which is connected to both Alice and Bob via the optical fiber links. Using decoy-states, a lower bound for the secret-key generation rate of $1.04 \times 10^{-6}$ bits/pulse is computed.

## 1. INTRODUCTION

The secure transfer of information is a necessity in the ever-connected modern world, and quantum key distribution (QKD) is a very promising idea that is able to theoretically achieve this objective [1]. QKD has been extensively experimentally demonstrated both in point-to-point links [2-6] and in quantum network configurations [7]. The intrinsic security of this technology is guaranteed by the principles of quantum physics, contrary to classical cryptography protocols that are based on mathematical complexity [1]. The security of QKD has been unconditionally proven [8], and even when imperfect devices are taken into account [9]. However, current technology presents loopholes that can be exploited by an eavesdropper to obtain information on the secret key [10-12]. Recently, a series of side-channel attacks have been proposed and demonstrated exploring imperfections on the single-photon detector (SPD) with possible countermeasures discussed [13-21]. To avoid that this type of backdoor remains open in a QKD system, a device-independent QKD protocol has been proposed [22]. However, the need for high detection efficiency in order to perform a loophole-free Bell inequality violation makes this proposal impractical from the point-of-view of currently available technology.

A new family of device-independent protocols has been recently proposed, which removes these so-called side-channel attacks [23,24]. Furthermore, the detectors may have been taken over by the eavesdropper without compromising the security of the system. While [24] makes use of entangled systems and a quantum memory for replacing real channels by secure virtual channels, the so-called measurement device independent (MDI)-QKD [23] is well suited to be used with practical weak coherent pulses, making use of the decoy states method. This is similar to a time-reversed EPR-based protocol, with two remote stations (Alice and Bob) sending faint laser pulses with individually randomly encoded quantum states to be measured at a mid-way station (Charlie), which contains a Bell-state analyzer (BSA) [23]. The BSA is used to project the two-photon states of both incoming optical pulses onto Bell states. Even if weak coherent states are used, the single-photon contribution is retrieved thanks to the usage of decoy states. Charlie broadcasts the measurement outcomes, which are used by Alice and Bob to agree on a shared secret key. As envisioned in the original proposal [23], the MDI-QKD removes all detection side-channels, as the eavesdropper may even impersonate the third party, "Charlie", without jeopardizing the security of the distributed key. MDI-QKD has already drawn a great interest from the experimental QKD community, with two recent demonstrations, both employing time-bin qubits [25,26].

In this paper we perform an experimental implementation of MDI-QKD using polarization qubits encoded on weak coherent states, as originally proposed by Lo *et al* [23]. The links between Alice (Bob) and Charlie are 8.5 km optical fiber spools, individually stabilized with automatic polarization control systems [27-30]. This is fundamental to guarantee a good fidelity between the chosen SOPs of the faint laser pulses sent by Alice and Bob, and the arriving SOPs after fiber transmission at the BSA. The polarization stability of both fibers was previously assessed by monitoring the Hong-Ou-Mandel (HOM) [31] interference visibility of fully independent weak coherent states over time [30]. The different valid combinations of SOPs sent by Alice and Bob, i.e. under bases agreement, are analyzed over the fiber-based remote BSA, also previously tested with the automatic polarization control [32]. We then employ the decoy state method with vacuum + weak state for MDI-QKD [33-38], allowing us to obtain a lower bound for the final secure key rate.

## 2. MDI-QKD

The MDI-QKD protocol [23] requires that Alice and Bob independently produce single photons, in the form of indistinguishable weak coherent pulses (WCPs), which are sent to Charlie and projected onto one of the four Bell states. Fig. 1 shows a simplified setup with a partial BSA.

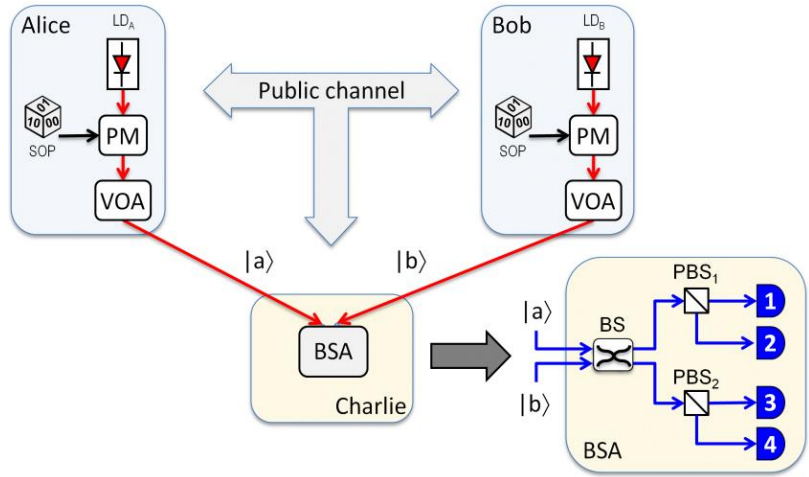

Figure 1. MDI-QKD scheme with the partial Bell state analyzer (BSA) based on linear optics.

The SOPs of the faint pulses are randomly and independently encoded in a state from one out of two mutually unbiased bases in a two-dimensional Hilbert space. The rectilinear basis (⊕), comprised by horizontal and vertical SOPs, and the diagonal basis (⊗), defined by the SOPs +45° and -45°, are usually chosen, similarly to traditional QKD protocols such as BB84 [1]. The output of Alice and Bob's laser diodes (LD) are therefore randomly polarization-modulated, according to the basis and bit choice, and sent to Charlie, after passing through variable optical attenuators (VOA) to set the desired average photon number per pulse. A polarization BSA based on linear optics performs the measurements on the incoming photons, and the results are publicly reported to Alice and Bob, which then proceed on to basis reconciliation, error correction and privacy amplification, as in traditional QKD protocols [2].

The possible results from the BSA are single detections in any of the four detectors, due to one or more photons routed to the same SPD (which are inconclusive events), and coincidence counts between pairs of single-photon detectors. All the single detections are ignored by the protocol, since currently available APD-based SPDs do not usually present photon number resolution to discriminate them from single-photon events (note that in some time windows a photon number combination different than one from both Alice and Bob may reach the BSA). The coincidence events are publicly announced by Charlie as conclusive BSA outcomes. In the basis reconciliation step, all instances in which Alice and Bob have each sent a state from incompatible bases ($\oplus$ with $\otimes$ and vice-versa) are also discarded, just like in traditional BB84-based protocols. The basis-compatible valid results from the BSA are summarized in Table I below for the ideal case of single-photon Fock states simultaneously sent by Alice and Bob and also for weak coherent pulses. It is worth noting that the decoy states modification enables Alice and Bob to retrieve information regarding the single-photon pulse statistics when coherent states are used [23], which are, finally, the useful events for the key generation.

TABLE I. Probability response of Charlie's BSA under the MDI-QKD protocol for single-photon pulses and WCPs under bases agreement (all other combinations are discarded during basis reconciliation). Rectilinear basis is used for key generation, while the diagonal basis is used for security purposes. Only the useful coincidences relating to the Bell state projections in states $|\psi^+\rangle$ or $|\psi^-\rangle$ have been included here [23].

| $\oplus$ basis | | BSA output | | | | $\otimes$ basis | | BSA output | | | |
|---|---|---|---|---|---|---|---|---|---|---|---|
| SOPs | | Single photons | | WCPs | | SOPs | | Single photons | | WCPs | |
| Alice | Bob | $|\psi^+\rangle$ | $|\psi^-\rangle$ | $|\psi^+\rangle$ | $|\psi^-\rangle$ | Alice | Bob | $|\psi^+\rangle$ | $|\psi^-\rangle$ | $|\psi^+\rangle$ | $|\psi^-\rangle$ |
| $|H\rangle$ | $|H\rangle$ | 0 | 0 | 0 | 0 | $|+45\rangle$ | $|+45\rangle$ | 1 | 0 | 0.75 | 0.25 |
| $|V\rangle$ | $|V\rangle$ | 0 | 0 | 0 | 0 | $|-45\rangle$ | $|-45\rangle$ | 1 | 0 | 0.75 | 0.25 |
| $|H\rangle$ | $|V\rangle$ | 0.5 | 0.5 | 0.5 | 0.5 | $|+45\rangle$ | $|-45\rangle$ | 0 | 1 | 0.25 | 0.75 |
| $|V\rangle$ | $|H\rangle$ | 0.5 | 0.5 | 0.5 | 0.5 | $|-45\rangle$ | $|+45\rangle$ | 0 | 1 | 0.25 | 0.75 |

Since a linear optics-based partial BSA is being employed ([32] and references within), only the $|\psi^+\rangle$ and $|\psi^-\rangle$ Bell states, which correspond to coincidence detections of the types $C_{12}$ or $C_{34}$, and $C_{14}$ or $C_{23}$, respectively, where $C_{ij}$ are coincidences between detectors $i$ and $j$ (Fig. 1), can be unambiguously distinguished from the other Bell states $|\phi^+\rangle$ and $|\phi^-\rangle$, with these last cases corresponding to two photons in a single spatial mode (same detector in Fig. 1). When both Alice and Bob send identical SOPs in the $\oplus$ basis, the BSA never outputs a valid coincidence detection event ($C_{13}$ and $C_{24}$ are not related to Bell projection) [23,32]. When orthogonal polarizations in the $\oplus$ basis arrive at Charlie, the BSA outputs $|\psi^+\rangle$ or $|\psi^-\rangle$ and either Alice or Bob need to perform a bit flip to correlate their bits. Finally, when the $\otimes$ basis is used, the BSA will output $|\psi^+\rangle$ or $|\psi^-\rangle$, depending on the combination of sent states according to Table I, and a bit flip is needed only when $|\psi^-\rangle$ is produced in the BSA. The bits produced from the $\oplus$ basis are used to form the shared secret key between Alice and Bob, whereas the other basis is used to test the error rate and channel transmittance by using decoy states [23]. This consists of randomly sending pulses with different mean photon numbers for testing against selective eavesdropping on multi-photon pulses, as the yields (probability that Bob detects a single-photon pulse) and quantum bit error rate (QBER) for single photon pulses can be extracted.

Since weak coherent pulses present a Poisson distribution for the photon number statistics, the probability of two photons being emitted by Alice while a vacuum pulse is sent by Bob (or *vice-versa*) is half the probability of a single-photon being emitted simultaneously by each party. This feature introduces a noisy offset to the diagonal basis, causing 25% of $|\psi^-\rangle$ events to occur when Alice and Bob's SOPs are equal (against 75% of $|\psi^+\rangle$) and 25% of $|\psi^+\rangle$ events when orthogonal states are sent (against 75% of $|\psi^-\rangle$). Additionally, coincidences $C_{13}$ and $C_{24}$ will occur, but are discarded as they play no role on the protocol. In spite of the spurious coincidence events, the protocol is still robust when implemented with WCPs because there is no impact on the rectilinear basis. Furthermore, the statistics of events generated from a pair of single photons in the diagonal basis can be extracted by applying the decoy states method [37,38], as originally proposed [23].

## 3. EXPERIMENTAL SETUP

In order to demonstrate the feasibility of the MDI-QKD protocol employing polarization encoding with current optical fiber technology, we performed a proof-of-principle experiment where Alice and Bob are each connected to Charlie through independent 8.5 km spooled fiber links. Due to the random residual birefringence fluctuations in the fibers we employed an active full-polarization control system based on two feedback signals wavelength-multiplexed with the single photons, thus stabilizing any polarization state transmitted through the fiber in the quantum channel wavelength, similarly to [27-30,32]. These two signals are generated from semiconductor distributed feedback (DFB) laser diodes located within Charlie's station as shown in the experimental setup (Fig. 2), and are split by 50:50 couplers, such that both signals are sent via the two fiber links in counter-propagating direction with respect to Alice and Bob's faint laser pulses. Both control signals are filtered using a combination of fiber Bragg gratings (FBGs) centered at the respective laser wavelength and an optical circulator, and detected in order to provide the feedback control signal to the automatic polarization control (APC) units.

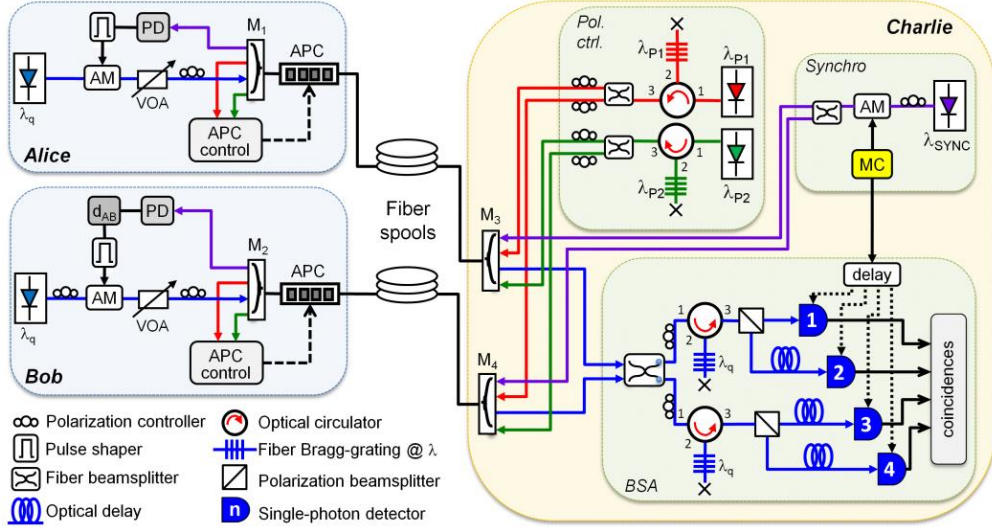

Figure 2. Experimental setup. AM: amplitude modulator, PD: photo-detector, VOA: variable optical attenuator, SOP: polarization controller to SOP preparation, M: WDM, APC: automatic polarization controller, MC: master clock, d: delay generator.

Alice and Bob have fully-independent continuous-wave (CW) tunable external cavity lasers, which together with $LiNbO_3$ amplitude modulators (AM), VOAs and manual polarization controllers (SOP, in Fig. 2) are used to produce the 1.5 ns wide polarization encoded faint pulses at 1546.12 nm. A polarimeter is used to verify the states of polarization. The VOA has a built-in mechanical shutter, which allows us to obtain a reliable estimation of gain of the vacuum state. The choice between signal and weak decoy states is performed by independently adjusting the VOAs attenuation values.

The temporal framework must be established previously to the communication. This means that the communicating parties must be able to identify and correlate their time slots. Previously to start key-distributing, Alice and Bob must adjust the relative delay between their signals relative to Charlie, which could be at different distances to Alice and to Bob (in the experiment, two 8.5 km long were used for convenience but asymmetric link lengths are allowed). Charlie also distributes the master clock signal (MC) to Alice and Bob via the same optical fibers by multiplexing in a full-duplex way another laser at $\lambda_{SYNC}$ = 1547.72 nm, which is pulsed using another amplitude modulator. The synchronism signal is demultiplexed and detected at Alice and Bob's stations with standard *pin* photodetectors (PD) and used to drive the AM that creates the weak coherent pulses. A delay generator ($d_{AB}$) is used at Bob's station to adjust and match the relative delay of the optical paths leading Alice and Bob to Charlie, assuring that both optical pulses arrive simultaneously at the BSA. Charlie must also ensure synchronism of his SPDs with Alice/Bob, which is implemented here with an electric delay generator (*delay* block, in Fig. 2) after the master clock.

The impact of the classical polarization control and the synchronization signals on the noise in the quantum channel is properly managed with the optical filters employed, as well as with appropriate optical input powers. All measurements reported in Section 4 – including coincidence counts when both Alice and Bob sources are turned off (vacuum-vacuum decoys) – were performed with all signals co-existing in the optical fibers.

We reinforce that the proposal of the MDI-QKD protocol is to avoid all detection loopholes, what means that any detector-related security issue is removed and the detection apparatus of the middle-way station (Charlie) may even be under the eavesdropper control [23]. In this context, the addition of the three auxiliary channels used in the experiment reported has no influence on security. The polarization control signals contain no information on the particular choice of bases by Alice and Bob. Any polarization state is stabilized by the system, regardless the particular choice of polarization state transmitted [27]. It is necessary to guarantee that Alice, Bob and Charlie bases are matched. There would be no information gain if an eavesdropper were to obtain information regarding the orientations of the reference states or the polarization bases, as it is always assumed Eve has a complete knowledge of it. Furthermore, the synchronism signal has no information regarding the encoding degree of freedom and only provide the temporal reference to synchronize Alice and Bob pulses at Charlie station.

The BSA is composed of a 50:50 fiber coupler, two polarizing beamsplitters (PBS) – preceded by polarization controllers – and the four SPDs. The master clock is delayed by twice the propagation time of the fiber in order to trigger each SPD synchronously to the optical pulses. Two additional circulator + FBG combinations centered at the faint pulses wavelength $\lambda_q$ are included to filter out any cross-talk photons which may have leaked in the DWDMs and minimize the impact of the classical signals on the quantum channel. The detection events are sent to a four-channel coincidence unity, which acquires all combinations of SPDs that fire during the same bit-slot. The SPDs used are InGaAs-based commercial detectors working in gated mode, with 2.5 ns gate windows, 1 MHz repetition frequency, average dark count probability per gate of $1.5 \times 10^{-5}$ and 10 $\mu s$ deadtime applied after each detection event. The BSA root-mean-square misalignment error was previously evaluated as 1.9%, and its temporal stability, including the automatic polarization control, was assessed under thermal induced birefringence on long optical fiber spools [32]. All system components outside the polarization controlled sections are sufficiently stable in an environment with regular air-conditioning system.

In order to correct small optical frequency drifts between both lasers, which are crucial for the correct operation of the protocol, an optical tap is taken at each laser´s output and recombined at a photodetector, thus monitoring the beat frequency between them (not shown in Fig. 2 for the sake of simplicity). The wavelength of one laser is automatically corrected through a feedback control loop if the drift exceeds 10 MHz. A solution for the practical case that Alice and Bob are spatially separated is through the use of a gas cell, such as HCN, at Alice and Bob's stations to independently lock each laser at the same spectral absorption line.

Temperature effects on the synchronization signal were not considered here and are naturally minimized by our setup configuration because both fibers have similar lengths in the same environment. Refractive index and length variations of fused silica with temperature are of the order of 10 ppm/K; this means that if the difference in length between Alice and Bob's fibers were of the order of a few kilometers, this would imply a temperature drift of arrival times of the order of a few tenths of a ns/K, still within the tolerance of the HOM dip. Hence, for laboratory purposes, synchronization is not a problem. However, if the system is to be installed in the field, a correction procedure to adjust the relative delay between Alice and Bob timing must be considered. This can be automatically performed, for example, if Charlie performs the delay adjustment ($d_{AB}$) instead of Bob, using the polarization control channels to separately trigger Alice's and Bob's pulses. Both immediately send towards Charlie a classical pulse through the synchronization channel, at orthogonal polarizations in order to be distinguished by Charlie, which adjusts $d_{AB}$ to compensate for any time-of-flight variations.

Moreover, it should be mentioned that the time interval between consecutive pulses from Alice and Bob (1 µs) is much greater than the coherence time of the photons, such that phase randomization is naturally obtained [39].

## 4. RESULTS AND DISCUSSIONS

The feasibility of the experiment depends on the experimental realization of the HOM interference effect between two independent laser sources with large spatial separation between them (8.5 + 8.5 km in our case), which is already shown in [30,32] for the CW case. Here, the independently pulsed WCPs were synchronized for temporal overlapping at Charlie's station and were made indistinguishable by overlapping their frequency and SOPs ($|H_a H_b\rangle$).

Bob's delay was scanned up to 4.5 ns far from the maximum temporal overlap and the coincidence count rate between $SPD_1$ and $SPD_3$ was recorded with Alice and Bob sending an average of 1 photon per 1.5 ns wide pulse. By shifting the wavelength by 0.02 nm, the coincident rate for distinguishable photons was also recorded. Interference visibility was calculated by $V = (C_d - C_i)/C_d$ [40], with $C_d$ and $C_i$ standing for distinguishable and indistinguishable coincidence rates, respectively, resulting in a coincidence dip (HOM-dip), depicted in Fig. 3.

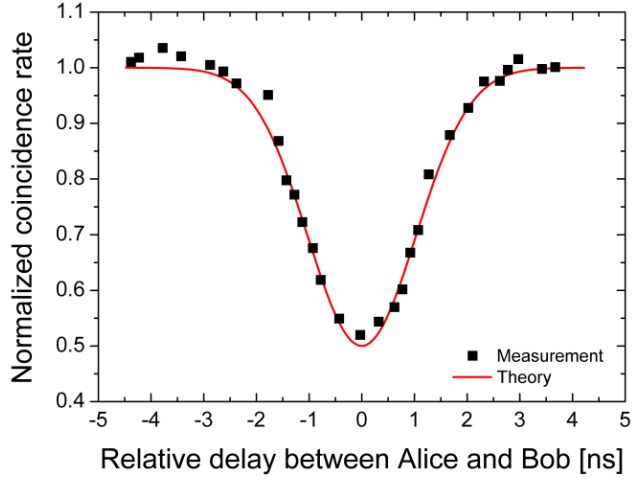

Figure 3. Visibility between the independent faint laser sources measured as a function of the relative optical pulses delay. For the visibility calculation, the values for distinguishable photons were obtained by detuning one laser by 0.02 nm.

The measured dip width is 2.40 ns. Close to maximum visibility is achieved, actually limited to 0.5 in the case of independent sources with Poisson distribution of the photon number statistics [40].

We performed an experimental simulation of a MDI-QKD session with decoy states. All measurements were performed with matched temporal modes, as required by the protocol. The automatic polarization control, previously assessed, was used to continuously compensate the random birefringence evolution of the quantum channel, by monitoring the WDM classical control channels.

The procedure described in [23] (see the supplemental material therein) was followed to achieve the gain and error values at each basis. The key parameters necessary for computation of the lower bound for the final key rate for the MDI-QKD protocol may be obtained by different means [23,37,38]. The weak + vacuum decoy states with linear programming approach [37] was chosen as it is a good approximation of the infinite decoy levels protocols while being much more practical and feasible.

In the protocol, Alice and Bob can randomly and independently choose between three different mean numbers of photons per pulse, representing different classes, the signal states, with $\mu_1 = 0.5$, and two decoy states, $\mu_2 = 0.1$ and $\mu_3 =$ vacuum. The signal average number of photons per pulse was chosen according to [37] and the weak decoy state level was chosen to be low but comfortably adjustable. Their SOPs were varied to cover the eight possible combinations with bases agreement, with all nine combinations of pulse classes, and the coincident counts were totalized during equal time intervals. When both classes are vacuum, the yield (probability of Charlie obtaining $|\psi^+\rangle$ or $|\psi^-\rangle$ given Alice and Bob sent a pulse) corresponds to the measured background count rate. For each pair of pulse classes at each pair of SOPs, the $C_{12} + C_{34}$ and the $C_{14} + C_{23}$ events, which correspond to the valid output from the relay station, respectively equivalent to $|\psi^+\rangle$ or $|\psi^-\rangle$ states, were recorded.

The first task is to prepare the gain ($Q_{i,j}^r$ and $Q_{i,j}^d$) values at each rectilinear (*r*) and diagonal (*d*) basis from the coincidence measurements, where *i* and *j* represents the source classes with different average number of photons per pulse used by Alice and Bob respectively. The channel gain is defined as the probability of successful detection on the BSA when a pulse with *i* and a pulse with *j* photons are sent by Alice and Bob, respectively. As the $|\phi^+\rangle$- and $|\phi^-\rangle$-like states are not detectable with our BSA, the probability of successful events at each pair of SOPs A and B correspond to the sum $C_{sum}^{AB} = C_{12}^{AB} + C_{23}^{AB} + C_{14}^{AB} + C_{23}^{AB}$ divided by the pulse repetition rate. Each gain value matrix can be calculated from the measured data as:

$$Q_{i,j}^r = \frac{1}{4}\left(C_{sum}^{HH} + C_{sum}^{VV} + C_{sum}^{HV} + C_{sum}^{VH}\right) \tag{1}$$

$$Q_{i,j}^d = \frac{1}{4}\left(C_{sum}^{++} + C_{sum}^{--} + C_{sum}^{+-} + C_{sum}^{-+}\right) \qquad (2)$$

where the indices H, V, + and − represent the SOPs |H⟩, |V⟩, |+45⟩ and |-45⟩ chosen by Alice and Bob.

Next we need to compute the error values, $E^{ij}_r$ and $E^{ij}_d$, corresponding to the abnormal occurrence of valid events on the relay measured at each basis for each pair of source classes. It is worth noting that, on the rectilinear basis, equal SOPs ideally give origin to no detection events on the BSA. So any coincidence is regarded as an error. However, when orthogonal states are used, all detection events are true detections and no error can be caused (any error would be converted to $|\phi^+\rangle$- or $|\phi^-\rangle$-like states, not detectable). So by the definition of QBER, i.e., the ratio between the number (or rate) of wrong events and the total number (or rate) of detected events, we can write

$$E_{i,j}^r = \frac{C_{sum}^{HH} + C_{sum}^{VV}}{C_{sum}^{HH} + C_{sum}^{VV} + C_{sum}^{HV} + C_{sum}^{VH}} \qquad (3)$$

Let us emphasize that the Poisson distribution of the number of photons of the sources has no practical influence here, because if Alice sends a pulse containing two photons (which have the same SOP), while Bob sends an empty pulse, the result will be a $|\phi^+\rangle$- or $|\phi^-\rangle$-like state (and the BSA will not detect), or a $C_{13}$ or $C_{24}$ event, which are not used on the protocol. However, the same does not hold for the diagonal basis. As shown in Table I, the Poisson distribution of the faint laser sources will cause a noise floor, which corresponds, mainly, to the cases in which Alice (Bob) sends a two-photon pulse while Bob (Alice) sends an empty pulse. In these cases, as any path can be taken by each photon at the BSA, erroneous $|\psi^+\rangle$- or $|\psi^-\rangle$-like states can occur, when a specific event is expected according to the states preparation. The QBER value for each pair of source classes in the diagonal basis can be calculated from measured data according to

$$E_{i,j}^d = \frac{C_{14}^{++} + C_{23}^{++} + C_{14}^{--} + C_{23}^{--} + C_{12}^{+-} + C_{34}^{+-} + C_{12}^{-+} + C_{34}^{-+}}{C_{sum}^{++} + C_{sum}^{--} + C_{sum}^{+-} + C_{sum}^{-+}} \qquad (4)$$

For weak coherent states the gain and error matrices can be written as [23]

$$Q_{i,j}^x = \sum_{n=0}^{N} \sum_{m=0}^{M} \frac{\mu_i^m \mu_j^n \exp(-\mu_i - \mu_j)}{m! n!} Y_x^{m,n} \qquad (5)$$

$$Q_{i,j}^x E_{i,j}^x = \sum_{n=0}^{N} \sum_{m=0}^{M} \frac{\mu_i^m \mu_j^n \exp(-\mu_i - \mu_j)}{m! n!} Y_x^{m,n} e_x^{m,n} \qquad (6)$$

with $x$ representing rectilinear or diagonal basis. These four equations (two for each basis) are used to extract the yield ($Y_x^{m,n}$) and QBER ($e_r^{m,n}$) values for each pair of Fock states $|m\rangle_{Alice}|n\rangle_{Bob}$ actually sent by Alice and Bob on both bases. The yield represents the conditional probability that two pulses with determined number of photons emitted by the optical sources reach Charlie. The Poisson distribution was expanded up to eight terms (M,N=7) for the linear programming [37] and the yield $Y_r^{11}$ is lower bounded, while the error $e_d^{11}$, which results in a lower bounded secret rate per pulse, given by [23,37]

$$R = Q_r^{11}\left[1 - H_2\left(e_d^{11}\right)\right] - Q_{rect} H_2(E_{rect}) f(E_{rect}), \qquad (7)$$

where $H_2(x)$ is the Shannon entropy, $Q_r^{11}$ is the channel gain for single photons pulses on rectilinear basis, $e_d^{11}$ is the QBER for single photon pulses at diagonal basis, $Q_{rect}$ is the global channel gain on rectilinear basis, and $E_{rect}$ is the global QBER on rectilinear basis.

The channel gain for pairs of pulses containing *m* and *n* photons exactly, sent by Alice and Bob respectively, on the rectilinear basis, is calculated from the values of yield ($Y_r^{m,n}$) and the probability of generation by the optical sources, i.e., $Q_r^{mn} = Y_r^{m,n} \mu_1^{m+n} \exp(-2\mu_1)/(m!n!)$, with the global gain in the rectilinear basis obtained by summing over the $Q_r^{mn}$ values. The global QBER for the rectilinear basis is calculated with $E_{rect} = \sum_m \sum_n Q_r^{m,n} e_r^{m,n} / Q_{rect}$. The amount of resources necessary on the error correction process is also discounted at the last term of Eq. (9), with *f(x)* standing for the inefficiency factor for the error correction [41].

The main values used for computing the secret key rate are reported in TABE 2, along with the gain and error values measured in the experiment. Using the calculated inefficiency factor for the error correction of 1.164, the secure key rate is computed as $1.04 \times 10^{-6}$ bits/pulse.

TABLE II. Gain and QBER values measured for each pair of source classes at each SOP basis. The values extracted with data analysis are also reported.

| WCPs | | Gain | | QBER | | Extracted data | |
|---|---|---|---|---|---|---|---|
| $\mu_i$ | $\mu_j$ | $Q_{i,j}^r$ | $Q_{i,j}^d$ | $E_{i,j}^r$ | $E_{i,j}^d$ | | |
| 0.5 | 0.5 | $9.44 \times 10^{-6}$ | $1.87 \times 10^{-5}$ | 0.057 | 0.296 | $Q_r^{11} =$ | $6.88 \times 10^{-6}$ |
| 0.5 | 0.1 | $2.19 \times 10^{-6}$ | $6.94 \times 10^{-6}$ | 0.093 | 0.393 | $E_d^{11} =$ | 0.018 |
| 0.5 | 0 | $3.96 \times 10^{-7}$ | $5.25 \times 10^{-6}$ | 0.463 | 0.479 | $Q_{rect} =$ | $1.36 \times 10^{-5}$ |
| 0.1 | 0.5 | $2.02 \times 10^{-6}$ | $6.65 \times 10^{-6}$ | 0.107 | 0.378 | $E_{rect} =$ | 0.057 |
| 0.1 | 0.1 | $6.25 \times 10^{-7}$ | $8.50 \times 10^{-7}$ | 0.060 | 0.240 | R= | $1.04 \times 10^{-6}$ |
| 0.1 | 0 | $4.17 \times 10^{-8}$ | $2.50 \times 10^{-7}$ | 0.400 | 0.417 | | |
| 0 | 0.5 | $3.08 \times 10^{-7}$ | $4.93 \times 10^{-6}$ | 0.378 | 0.496 | | |
| 0 | 0.1 | $4.17 \times 10^{-8}$ | $2.08 \times 10^{-7}$ | 0.300 | 0.400 | | |
| 0 | 0 | $4.90 \times 10^{-10}$ | $4.90 \times 10^{-10}$ | 0.500 | 0.500 | | |

The secure key rate was also calculated by using the MDI-QKD analytical model described in [37]. From the average dark counts rate, the evaluated misalignment error of the BSA and considering the measured channel loss of 19.5 dB per link, which includes the penalty imposed by the 15% detection efficiency of the SPADs, the calculated lower bound rate was $1.59 \times 10^{-6}$ bits per pulse, quite close to the value computed from the measured data.

The positive final secure rate proofs the principle of the MDI-QKD protocol and it is limited here by a number of technical details, most related to the channel gain. Replacing some optical components by state-of-the-art devices and by splicing all optical fiber connections outside Alice and Bob stations the channel transmittance could be increased by up to 4.3 dB. Furthermore, an improvement on the SPADs detection efficiency to 25% would reduce the channel loss by additional 2.2 dB. Considering such improvements, the calculated lower bound for the secret key rate would reach $4.78 \times 10^{-5}$ bits/pulse for the same link length. Conversely, without including any other issue, the $>1.04 \times 10^{-6}$ bits/pulse secret key rate could be established between Alice and Bob up to 82 km apart, with Charlie at midway.

## 5. CONCLUSIONS

We performed a proof-of-principle demonstration of a polarization-encoded measurement-device-independent quantum key distribution (MDI-QKD) protocol with weak coherent states transmitted over two optical fiber links of 8.5 km each. Each fiber link connecting Alice and Bob to Charlie's station was independently stabilized against polarization drifts using a full polarization control system employing two wavelength-multiplexed control channels. One additional channel was used to synchronize Alice and Bob's optical pulses using the same optical fibers in a counter-propagating direction. The decoy-states protocol was carried out employing signal pulses with 0.5 photon on average and decoy pulses with 0.1 and 0 (vacuum) photons. The lower bound for the secure key generation rate was $1.04\times 10^{-6}$ bits/pulse, showing that polarization encoded MDI-QKD is feasible in long optical fibers. Furthermore, our polarization control system is compatible with classical high-speed telecom optical data channels, further enhancing the possibility of implementing this experimental setup in a fiber-optics telecom environment.

ACKNOWLEDGMENTS

Authors would like to thank X.-B. Wang and X. Ma for enlightening discussions. This work was supported by Brazilian agencies CAPES, CNPq and FAPERJ. In addition G. B. Xavier acknowledges support from grants Milenio P10-030-F CONICYT PFB08-024 and FONDECYT 11110115.